# Direct imaging of local atomic structures in zeolite using novel low-dose scanning transmission electron microscopy

Kousuke Ooe[1,3], Takehito Seki[1,2*], Kaname Yoshida[3], Yuji Kohno[4], Yuichi Ikuhara[1,3], Naoya Shibata[1,3*]

[1]Institute of Engineering Innovation, School of Engineering, the University of Tokyo, Yayoi 2-11-16, Bunkyo, Tokyo, 113-0032, Japan.
[2]PRESTO, Japan Science and Technology Agency, Kawaguchi, Saitama, 332-0012, Japan
[3]Nanostructures Research Laboratory, Japan Fine Ceramics Center, Mutsuno 2-4-1, Atsuta, Nagoya, 456-8587, Japan.
[4]JEOL Ltd., 1-2-3 Musashino, Akishima, Tokyo 196-8558, Japan.

[*]Corresponding authors.
Email: seki@sigma.t.u-tokyo.ac.jp  (T.S.); shibata@sigma.t.u-tokyo.ac.jp (N.S.)

## Abstract

Zeolites have been used in industrial applications such as catalysts, ion exchangers, and molecular sieves because of their unique porous atomic structures. However, the direct observation of zeolitic local atomic structures via electron microscopy is difficult owing to their low resistance to electron irradiation. Subsequently, the fundamental relationships between these structures and their properties remain unclear. A novel low-electron-dose imaging technique, optimum bright-field scanning transmission electron microscopy (OBF STEM) has recently been developed. It reconstructs images with a high signal-to-noise ratio and a dose efficiency approximately two orders of magnitude higher than that of conventional methods. Herein, we performed low-dose atomic-resolution OBF STEM observations of an FAU-type zeolite, effectively visualizing all the atomic sites in its framework. Additionally, the complex local atomic structure of the twin boundaries in the zeolite was directly characterized. The results of this study facilitate the characterization of the local atomic structures in many electron-beam-sensitive materials.

# Introduction

Zeolites are porous materials with regularly arranged nanosized pores, which enable a wide range of applications in catalysis, gas separation, and ion exchange [1,2]. The material properties of zeolites are closely related to the geometry of their pores and their subsequent interactions with any adsorbed guest molecules and ions. To date, diffractometric techniques have been most often used for the structural analysis of zeolites [3]. Although diffraction methods can accurately analyze averaged structures, obtaining local structural information related to defects, interfaces, and surfaces is extremely difficult. Scanning transmission electron microscopy (STEM) is a powerful technique for local structural analysis that enables the direct observation of atomic structures in electron-resistant materials at a sub-angstrom resolution [4]. However, zeolites are more electron beam-sensitive than other inorganic materials. Thus, atomic-scale observations via electron microscopy are severely limited by electron irradiation damage [5,6]. Menter observed faujasite zeolite for the first time in 1958 via high-resolution transmission electron microscopy (HRTEM), and reported a lattice resolution of 14 Å [7]. Subsequently, the zeolite framework structure was observed [8]. In the 1990s, an aberration corrector was developed and the S/TEM resolution was significantly improved [9]. These technological advances have enabled the direct observation of the framework structure and arrangement of adsorbed cations in zeolites [10]. However, it remains extremely challenging to directly observe all the atomic sites in zeolites, including the Si/Al and oxygen sites, owing to the severe electron irradiation damage within zeolites.

Recently, the development of new STEM electron detectors has led to more advanced imaging techniques. Although conventional STEM uses a single annular detector to detect transmitted/scattered electrons to form images, the recently developed segmented/pixelated detectors can simultaneously form many STEM images using electrons detected in multiple areas on the diffraction plane. By processing these multiple STEM images, information regarding the electromagnetic fields and phase information of the samples can be obtained [11,12]. We have theoretically developed an optimum bright-field (OBF) STEM technique for low-dose imaging that enables the observation of atomic structures at the highest signal-to-noise ratio (SNR) using segmented/pixelated detectors [13]. Fig. 1a shows the schematic of the OBF STEM technique, which uses a segmented detector. Herein, a finely focused electron probe is raster-scanned across the sample, and the transmitted/scattered electrons are detected at each raster by a multiple-segmented electron detector. Subsequently, frequency filters are applied to each image obtained by the corresponding detector segment, and the filtered images are assembled to obtain the OBF image. These filters were designed to maximize the SNR of the synthesized image based on the STEM contrast transfer function (CTF) [14] and noise-evaluation theory [15]. Fig. 1b presents the noise-normalized CTFs for various phase-contrast STEM techniques. The noise-normalized CTF represents the SNR as a function of spatial frequency and is helpful in evaluating the imaging efficiency of different techniques [15]. OBF STEM, which uses a segmented detector, exhibits a much higher imaging efficiency than those of conventional techniques such as annular bright-field (ABF) and conventional bright-field (BF) imaging [16] over an entire spatial frequency domain. Additionally, OBF is more efficient than integrated differential phase-contrast (iDPC) imaging [17], which is a phase-imaging technique that also uses a segmented detector. As described in Supplementary Text, OBF imaging achieves a dose efficiency approximately two orders of magnitude higher than those of the conventional STEM imaging methods and is ~24% higher than that of iDPC imaging. Furthermore, OBF can obtain information at much higher spatial frequencies (i.e., higher resolution) than those of the conventional techniques, as shown in Fig. 1b. In this study, we obtained atomic-level structural images of zeolite via aberration-corrected STEM with an accelerating voltage of 300 kV and a probe-forming aperture of 15 mrad, wherein the information limit of the OBF contrast transfer was calculated to be 0.66 Å in real space, indicating the present optical condition to be sufficient for obtaining images with atomic resolution.

Furthermore, the OBF images could be reconstructed in real time [13]. In a typical atomic-resolution STEM operation, fine optical tuning adjustments, such as astigmatism correction, defocus correction, and field-of-view (FOV) adjustment, are performed by an operator who refers to atomic-resolution images displayed on a monitor in real time. In low-dose observations, fine tuning becomes much more difficult because the operator cannot observe atomic structures in the real-time images owing to poor SNR. However, the photomultiplier-based segmented detector enables the dwell time of the electron probe to be as short as that of conventional detectors, and the synthesized images can also be processed as live imaging [18]. Thus, by implementing a real-time OBF imaging function combined with a high-speed segmented detector and rapid scanning, an operator can observe atomic structures in real time with a higher SNR and tune the optical parameters even under low-dose conditions, as shown in Movie S1. This technique facilitates the observation of beam-sensitive materials with minimal irradiation damage.

In this study, we used real-time OBF imaging to observe FAU-type zeolites with sub-angstrom resolutions. We demonstrated that OBF imaging allows the direct observation of the T (=Si, Al) and oxygen atoms in the $TO_4$ tetrahedron building units, which constitute the FAU-type framework structure. Furthermore, OBF imaging was used to directly observe the detailed atomic structure of a twin boundary, which is a common lattice defect in FAU zeolites. The results of this study highlight the capability of electron microscopy for the local structural characterization of beam-sensitive materials.

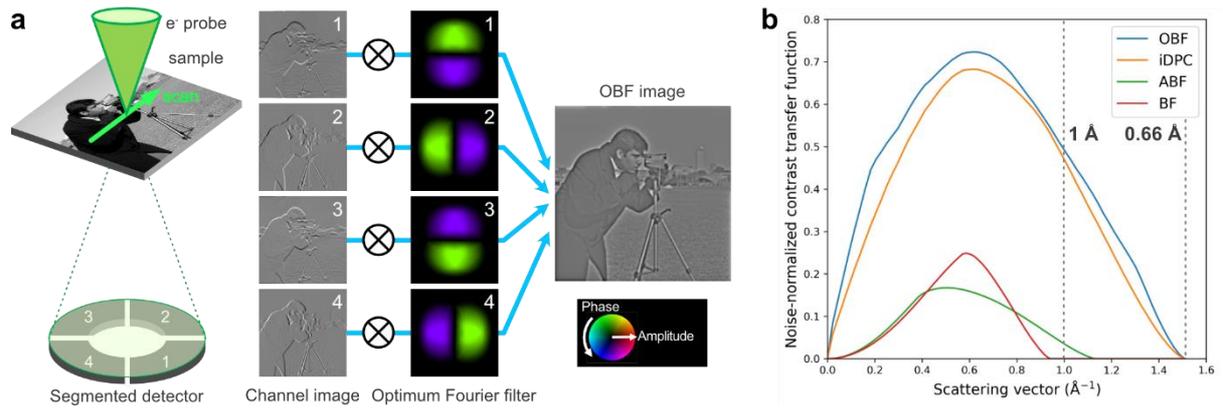

**Fig. 1. Reconstruction scheme of OBF STEM and dose-efficiency comparison based on noise-normalized CTFs for different STEM imaging techniques.** (a) Schematic illustration of OBF STEM image processing workflow. In OBF STEM, a segmented detector is located on the diffraction plane that collects the intensity of transmitted/diffracted electrons at each probe position. The STEM images acquired by each segment are then processed with frequency filters to extract the phase-contrast component. The frequency filters are derived via STEM CTF, which are of a complex value. Subsequently, the filters are also complex-valued and visualized as a color map representing the phase and amplitude. After filtering, all the images are summed, and the OBF image is synthesized. As the filter is calculated via microscope optical information such as accelerating voltage and convergence angle of the probe as well as the CTF, OBF reconstruction does not need *a priori* knowledge of the sample. (b) Noise-normalized CTFs of OBF and various phase-contrast imaging techniques. CTFs show the window of contrast transfer from samples as a function of spatial frequency. Noise-normalized CTF is calculated by normalizing CTFs based on the noise level at each spatial frequency within the Poisson statistics, which shows the SNR at each Fourier component. Herein, the CTFs are calculated at an accelerating voltage of 300 kV, a convergence semi-angle of 15 mrad, and a sample thickness of 10 nm: the same conditions as those of the experiments conducted in this study. These CTFs are shown as radially-averaged values, and the OBF technique shows a higher noise-normalized CTF than both the conventional methods (BF and ABF) and iDPC, the recently developed phase imaging technique.

# Results

## Direct imaging of atomic structures in FAU zeolite

Fig. 2a schematically shows the FAU framework, which consists of two building blocks: sodalite cages and double 6-membered rings (D6Rs). These two building blocks are connected with the same symmetry as that in a diamond structure and form large pores of 12 Å diameter. Implementing the real-time OBF imaging technique, we observed the FAU framework along the <110> zone axis, wherein the pores were aligned along the observation direction. Conventionally, the atomic structure from the same zone-axis was observed via HRTEM [19], but only the pore arrangements were resolved by this method, and the atomic resolution was difficult to achieve. Herein, the electron probe current was set to 0.5 pA to suppress the beam damage, which was approximately two orders of magnitude lower than that of the usual STEM observation condition for analyzing typical inorganic materials. Under these conditions, OBF STEM observations of the FAU-type zeolite sample were conducted, as shown in Fig. 2b-2e. Fig. 2b shows the experimental OBF image of the FAU-type zeolite. The OBF image of the FAU framework structure indicated the atomic sites as bright spots, evidently for the tetrahedral (T-sites occupied by Si or Al) and oxygen sites. An amorphous layer covering the sample surface [20] is also recognizable in the image. Fig. 2c shows the power spectrum of the OBF image in Fig. 2b that exhibits an information transfer up to 0.869 Å. Furthermore, Fig. 2d shows a unit cell-averaged OBF image obtained from the original OBF image in Fig. 2b. The FAU framework structure can be observed at an atomic resolution and conforms extremely well with the simulated image shown in the inset, indicating that the atomic structure can be resolved without any electron irradiation damage. Fig. 2e is the cropped image of Fig. 2d, focusing on the D6R building block of the FAU structure. It evidently shows that the tetrahedral units are connected via corner-shared oxygen sites. In zeolites, the oxygen-bridging sites between the tetrahedral units play an essential role in dictating the properties exhibited by the material, such as the catalytic activity [21] and structural transformation introduced by the interactions between the framework host and captured guests [22]. Thus, the capability of OBF STEM to visualize individual oxygen atom sites significantly helps to understand the structure-property relationship in zeolites. It is noteworthy that the visibility of the oxygen atom was already attained in the raw OBF image before unit-cell averaging. The SNR of the S/TEM images of zeolites is usually enhanced by averaging the raw data using *a priori* knowledge about the sample, such as the space groups of the material [23,24]. Although this is effective for homogeneous bulk structure analyses, it cannot be applied to heterogeneous or nonperiodic local structure analyses. Therefore, the presented direct atom imaging capability will be helpful for zeolitic heterogeneous/nonperiodic structure analyses, such as aluminum substitution, counter cations, and other defects. Later in this study, we have demonstrated the OBF STEM imaging of a defect structure in the FAU-type zeolite.

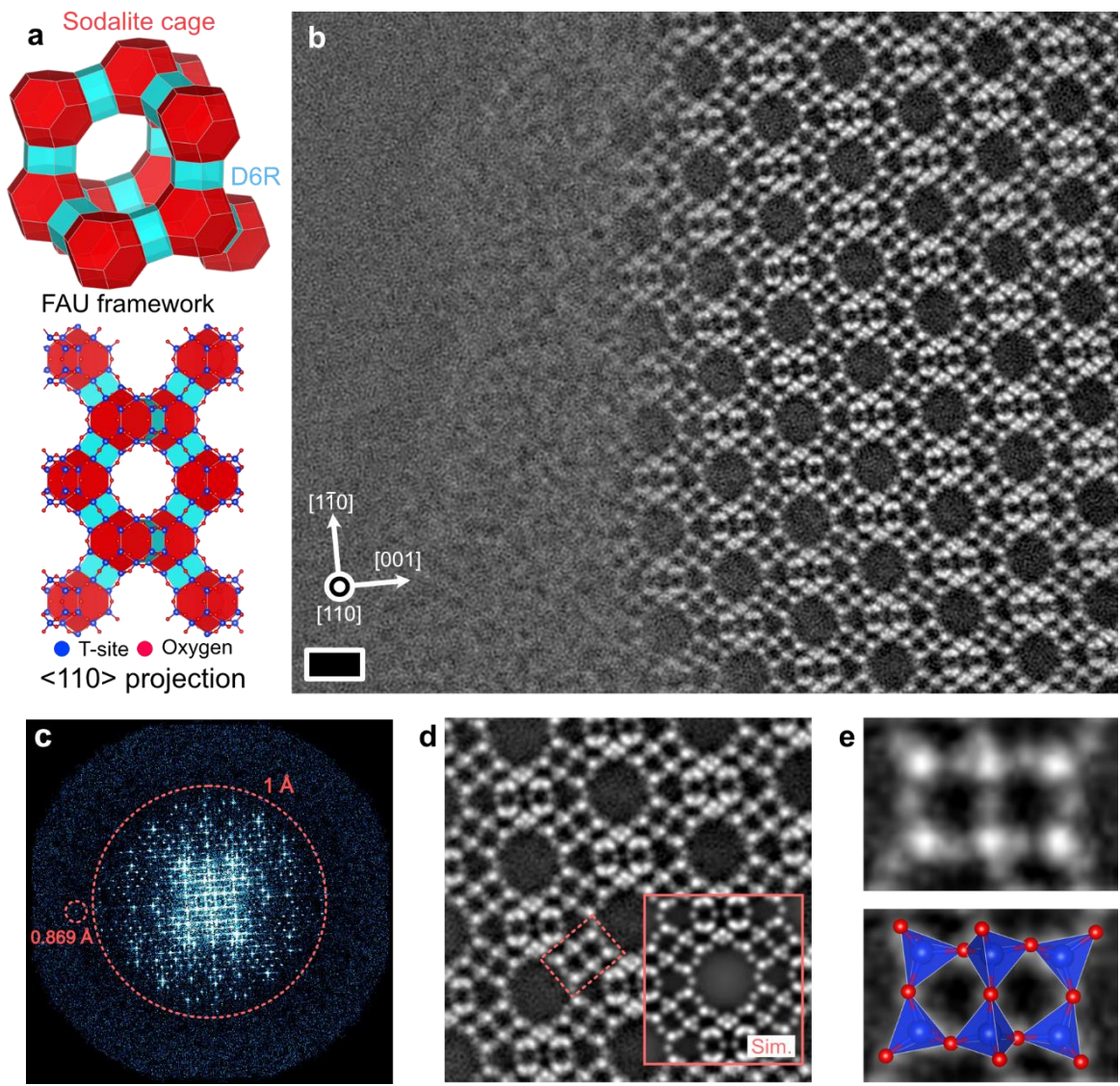

**Fig. 2. Atomic-resolution OBF STEM observation of an FAU zeolite along <110> zone axis.** (a) Schematic of the FAU zeolite framework structure and projected atomic-structure model along <110> zone-axis. Red and blue polygons represent the building units (sodalite cages and double 6-rings (D6Rs), respectively). (b) OBF STEM image of FAU zeolite observed at the edge of the sample. Bright spots indicate T- and oxygen-sites (scale bar: 1 nm). (c) Fast Fourier Transform (FFT) of (b), wherein FFT spots are seen up to 0.869 Å resolution in real space. (d) Repeat-unit-cell averaged OBF image. The inset is a simulated OBF image calculated with the same observation condition as that in the experiment. The location of the D6R structure, which is shown in (e), is highlighted by a dashed rectangle. (e) Magnified OBF image of the rectangular region indicated by the red dashed line in (d). The atomic structure models are drawn using VESTA [25].

We compared the OBF images with other STEM images obtained under the same dose conditions. Fig. 3a shows the OBF, iDPC, conventional BF, and ABF images. The OBF image shows the FAU framework structure with the highest SNR, conforming with the noise-normalized CTF calculation shown in Fig. 1b. Although the iDPC image also reveals the basic FAU framework structure, individual atomic sites, such as oxygen columns, are not distinguishable, as shown in the inset. For the further analysis of these contrast characteristics, we simulated the noise components of OBF and iDPC images (see the Materials and Methods section for details). In the OBF reconstruction, the noise level is set to be flat as a function of spatial frequency, which is known as the noise-normalized condition. This is equivalent to the so-called white noise. The noise fluctuation of the OBF image contrast is thus uniformly random for the entire FOV. It is noteworthy that this noise-component image is displayed on the same spatial scale as that of the experimental images shown in Fig. 3a for comparison. However, for the iDPC image, the contrast fluctuation due to noise exists on a spatially larger scale than that of the OBF image. This is confirmed by a line profile of the noise component. In the integration process of the DPC signal to form the iDPC image, the low-spatial-frequency component of the image is much more amplified than the higher-spatial-frequency components [26]. However, the iDPC signals essentially do not exhibit contrast transfer around the low-frequency domains against noise, as shown in Fig. 1b. Thus, under the low-dose condition, this amplification effect enhances the noise component in the lower frequency regions, resulting in the appearance of long-range contrast fluctuation, as shown in Fig. 3b. This is the reason why the iDPC image contrast appears smoother but has longer-range noise-fluctuation than those of the other methods. We also examined the experimental image intensity distribution of each imaging technique, as shown in Fig. S1, wherein the longer-range noise effect was more severe owing to the wide FOV. Although the OBF image exhibits an interpretable image contrast corresponding to the sample thickness and atomic sites, the iDPC image exhibits long-range intensity fluctuations in the experiment as well as the simulations, as shown in Fig. 3b. This contrast is much stronger than that of each atomic site in the zeolitic framework. This results in a poor visibility of the atomic sites and makes it difficult to interpret the atomic structures from the obtained image. In other STEM images, such as ABF and BF, the basic structure of the FAU framework is roughly visible, but the detailed atomic structure analysis is challenging under the present low-dose condition.

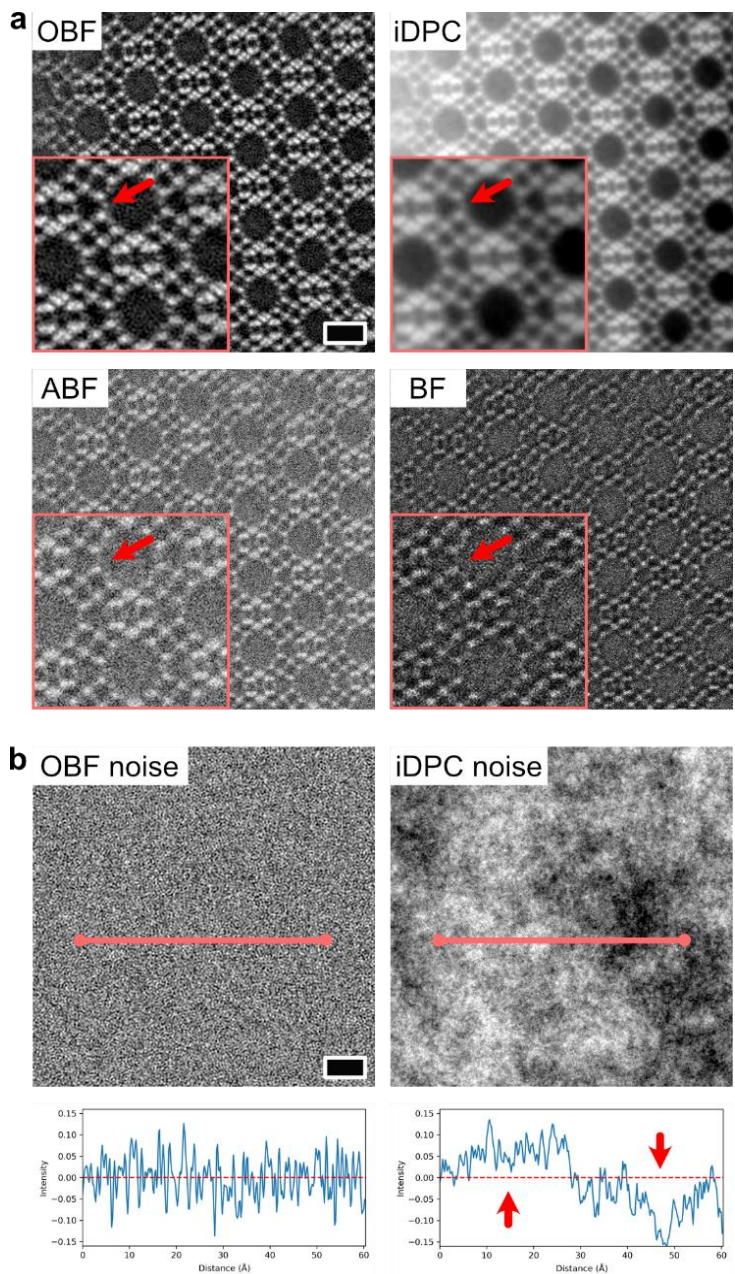

**Fig. 3. Comparison between atomic resolution images of OBF STEM and other STEM techniques.** (a) STEM images obtained via OBF, iDPC, ABF, and conventional BF imaging techniques (scale bar: 1 nm). All the images were recorded under the same electron dose and optical conditions (except for the defocus) as described in the Materials and Methods section. The insets are the cropped and enlarged versions of the original images, and the orange arrows indicate the oxygen sites in the FAU zeolitic structure. (b) Comparison of the noise components between the OBF and iDPC simulated images (see the Materials and Methods section for details). The intensity profiles of noise are also shown (obtained from the orange lines). The assumed dose is the same as that in the experiments shown in (a), and the noise components in both the methods are obtained by the same noise-introduced segmented-detector datasets. As indicated by the orange arrows, the iDPC noise image has longer-range fluctuations than that of the OBF.

**Direct observation of FAU twin boundary**

We applied the OBF technique to characterize the atomic structure of a twin boundary in the FAU zeolite. In FAU-type zeolites, the framework is constructed by cubic stacking of a layered structure unit called a 'faujasite sheet' [27]. When the faujasite sheets are stacked in a hexagonal sequence, the resultant framework exhibits an EMT-type structure, known as a polymorph of an FAU-type zeolite. There are twin boundaries between two opposite sequences of the cubic stacking in the FAU framework that likely result in an EMT-type structure at the boundary, as schematically shown in Fig. 4a. However, the detailed atomic structure could not be directly determined owing to the limited spatial resolution under the low-dose condition in the previous TEM study [28].

Fig. 4b shows the OBF image of the FAU twin boundary. This image indicates that the FAU cubic stacking sequence is inverted at the twin boundary. The power spectrum of the image indicates an information transfer beyond 1 Å. For further analysis, we averaged the structural units of the FAU twin boundary, as shown in Fig. 5a. The T and oxygen atomic sites are evidently visible in the twin boundary core, and two FAU-type domains are connected coherently at the atomic scale. Furthermore, the atomic structure of the twin boundary is confirmed to be identical to that of the EMT-type structure.

Density functional theory (DFT) calculations were performed to evaluate the stability of the twin boundary structure. The initial twin-atomic structure was created by stacking the faujasite sheets based on the OBF image and then relaxed via DFT calculations. Fig. 5c shows the relaxed atomic structure model, and Fig. 5b shows its corresponding simulated OBF image. The experimental image conforms well with its simulated counterpart. Furthermore, the interface energy was calculated to be 7.4 mJ/m$^2$, which is comparable with those of the twin boundaries in face-centered cubic (FCC) metals on the {111} plane [29], but approximately three orders of magnitude lower than those (typically) in oxide ceramic materials, such as grain boundaries and twin boundaries [30,31]. The origin of this difference can be explained as follows: in the {111} twin boundary of cubic zirconia, for example, the origin of the higher interface energy is attributed to the different coordination numbers of anions around the cation sites on the interface whereas the cation sites produce a coherent interface structure similar to those of FCC metals [32]. In the case of zeolites, the framework is constructed by the corner-sharing of rigid TO$_4$ tetrahedra, which have a nearly perfect tetrahedral shape and are connected via oxygen atoms as soft hinges, offering a rigid but stress-free atomic structure [33]. Thus, zeolites can relax their framework structure by simply changing the bond angle between two rigid TO$_4$ tetrahedrons (T-O-T angle). In silicate materials, the atomic structure is energetically stable over a wide range of T-O-T angles [34]. The observed structure of the FAU twin boundary was constructed in a similar manner, keeping the coordination numbers of the cations and anions unchanged across the boundary. This structural flexibility should result in extremely low excess energy at the twin boundary. Structural information about minute strains around some defects is essential for applications such as molecular sieves and gas separators. It may affect the diffusion process of ions and molecules adsorbed in the zeolitic nanocavity.

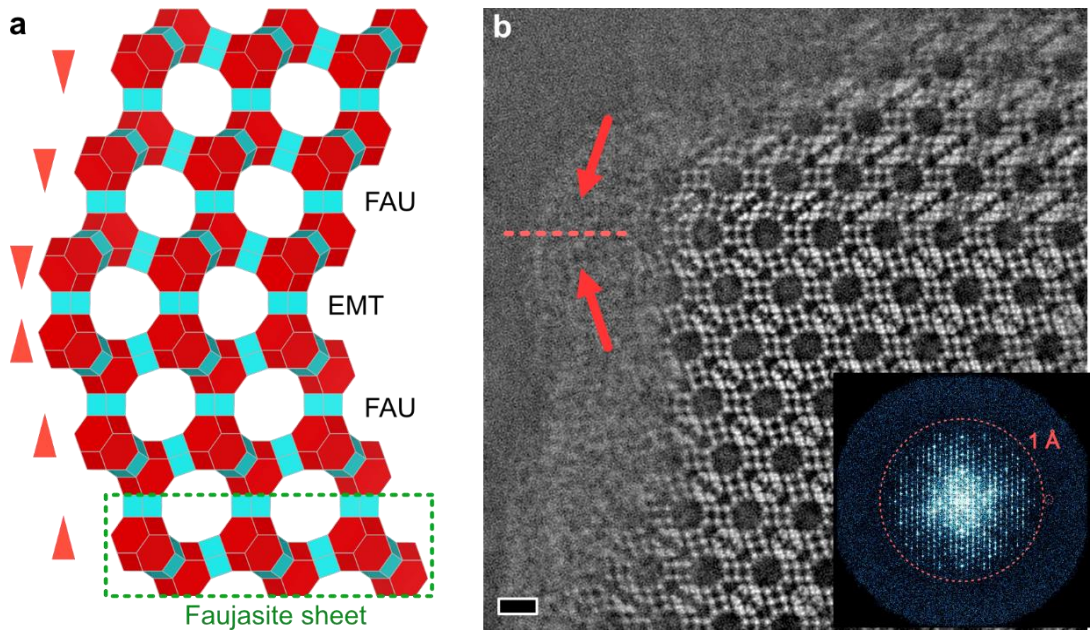

**Fig. 4. Atomic-resolution OBF STEM image of FAU twin boundary.** (a) The framework model of the FAU twin boundary. The FAU cubic stacking sequence is inverted on the twin boundary, making the EMT framework structure with hexagonal stacking. The structure highlighted with a green-dotted box is a faujasite sheet, a layer structure unit for the FAU and the EMT frameworks. The triangles represent the directions of the stacking sequence. (b) Atomic-resolution OBF STEM image of the FAU twin boundary (scale bar = 1 nm). The inset is the FFT pattern of the OBF image, which exhibits a contrast transfer beyond 1 Å.

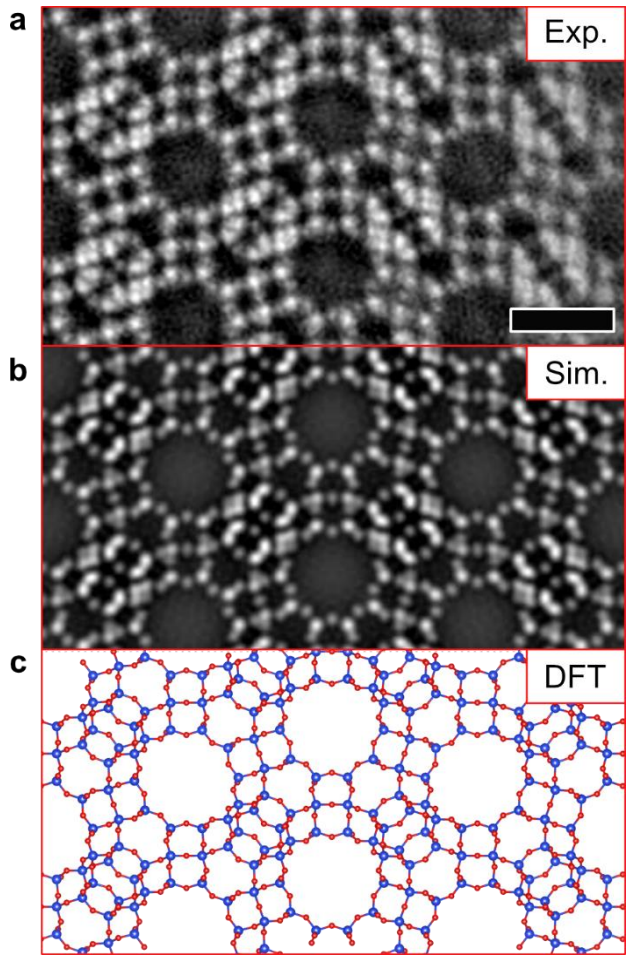

**Fig. 5. Comparison between experimental OBF image and simulated image based on DFT-relaxed structure of the FAU twin boundary.** (a) Unit-cell averaged experimental OBF image obtained from the raw experimental image shown in Fig. 4(b). The averaging operation is performed along the direction parallel to the interface, and no structural information is assumed about the symmetry other than the translational symmetry along the boundary (scale bar: 1 nm). (b) Simulated OBF image based on the DFT-relaxed structure shown in (c). The image is calculated under the same condition as that of the experiment. (c) Atomic structure model of FAU twin boundary relaxed by the DFT calculation. The blue and red balls indicate the T- and oxygen-sites, respectively. These images/structures show good agreement in both T- and oxygen-sites.

## Discussion

We developed a highly dose-efficient STEM imaging technique, OBF STEM, for application in low-dose atomic-resolution imaging. We demonstrated that OBF STEM can directly reveal the atomic structures of all elements in an FAU-type zeolite, which is a beam-sensitive material, with a sub-angstrom spatial resolution. OBF STEM can also be used to observe the lattice defects in zeolitic framework structures. We succeeded in directly determining the atomic structure on an FAU twin boundary, and the corresponding result was consistent with the DFT calculations. The proposed technique can thus be used to characterize the local atomic structure in zeolites and other beam-sensitive materials, facilitating the study of structure-property relationships in these materials.

## Materials and Methods

### Atomic-resolution OBF STEM observation of an FAU-type zeolite

Atomic-resolution OBF STEM images were acquired using an aberration-corrected STEM (JEOL JEM ARM-300F) equipped with a second-generation segmented annular all-field (SAAF) detector (16-segmented type) [35]. We developed an in-house program for the real-time OBF display function and implemented it in the SAAF system, as shown in Fig. S3. Movie S1 shows the real-time observation of a $SrTiO_3$ [001] sample with an accelerating voltage of 300 kV and a probe-forming aperture of 30 mrad. All the atomic columns, including the oxygen atoms, were visualized under a low-dose condition (probe current: 0.5 pA, i.e., two orders of magnitude less than the usual condition). This result demonstrated the capability of OBF STEM for low-dose and live atomic-resolution imaging. We used a real-time OBF display system to acquire all the experimental OBF images shown in the present study.

For the TEM sample preparation of an FAU-type zeolite, a commercially available powder sample of FAU zeolite (Tosoh Corp., Si/Al=50) was gently crushed in an agate mortar with ethanol and dispersed onto a TEM microgrid. Before STEM observation, the sample was dehydrated overnight in the high vacuum environment of the TEM column to suppress the irradiation damage [8]. The accelerating voltage was set to 300 kV, which effectively reduces the irradiation damage in the zeolites [6,36]. The probe current and probe-forming aperture were 0.5 pA and 15 mrad, respectively. Images of the FAU bulk structure were sequentially acquired at a dwell time of 16 µs with 1024 × 1024 pixels in the same region to suppress irradiation damage and scan distortion. Under this condition, the total dose was $1.2 \times 10^3$ e$^-$/Å$^2$ per frame. For the FAU twin boundary observation, the dwell time was reduced to 10 µs to further suppress the image distortion, with the total dose was $7.5 \times 10^2$ e$^-$/Å$^2$ per frame. After the sequential image acquisition, the first five images were aligned and averaged for each data set. Furthermore, we obtained unit-cell-averaged images for a detailed structural analysis, as shown in Figs. 2d and 5a. It can be noted that *a priori* knowledge about the structure group symmetry of the atomic structure was not assumed for the image averaging except for the translational symmetry for both FAU bulk and twin boundary analyses.

To obtain the OBF images, the camera length was set such that the edge of the STEM direct beam disk coincided with the outermost edge of the SAAF detector. Under these conditions, the OBF image was obtained using Equation (1):

$$I_{\text{OBF}}(\boldsymbol{R}_\text{p}) = \mathcal{F}^{-1}\left[\sum_{j=1}^{16} I_j(\boldsymbol{Q}_\text{p})W_j(\boldsymbol{Q}_\text{p})\right] = \sum_{j=1}^{16} I_j(\boldsymbol{R}_\text{p}) \otimes w_j(\boldsymbol{R}_\text{p}), \tag{1}$$

where $I_{\text{OBF}}(\boldsymbol{R}_\text{p})$, $I_j(\boldsymbol{Q}_\text{p})$, $W_j(\boldsymbol{Q}_\text{p})$, and $w_j(\boldsymbol{R}_\text{p})$ are the OBF image intensity, Fourier transformed image acquired by the *j*-th segment $I_j(\boldsymbol{R}_\text{p})$, frequency filter calculated for the *j*-th segment, and point spread function obtained via the inverse Fourier transform of the frequency filter $W_j(\boldsymbol{Q}_\text{p})$, respectively.

The filtering process was performed by multiplying the frequency filter in the reciprocal space $\boldsymbol{Q}_\mathrm{p}$ or convolution with the point spread function in the real space $\boldsymbol{R}_\mathrm{p}$. The post-processed OBF image could be obtained via either procedure, and the real-time OBF imaging synchronized with the STEM scan was acquired using the approximated convolution process in real space [13]. For the focal condition to obtain the STEM images, it was reported that the OBF and DPC image contrast can be theoretically maximized upon focusing the electron probe on the mid-plane of the specimen [13,37]. In contrast, the ABF and BF images exhibited the highest contrast upon focusing the probe on the entrance surface [16]. Thus, we acquired the images under the optimal focal conditions for each technique. To obtain the experimental/simulated iDPC, BF, and ABF images, the segmented/annular detector images were synthesized using the SAAF detector datasets to reproduce the detector geometry dedicated to each method.

**Image simulations**

For the STEM image simulation, we used the MuSTEM package [38] based on the multi-slice model [39]. 16-segmented-detector images were calculated and processed using the OBF reconstruction algorithm to obtain the simulated OBF image. The effective source size was considered by convolution with a 2D Gaussian with a full-width-half-maximum of 0.6 Å. The sample thickness was assumed to be 10 nm, and the defocus $\Delta f$ was set to middle-focus condition, wherein the focal plane is located at the mid-plane inside the sample ($\Delta f = -5$ nm).

The STEM image simulation was also used for noise property analysis, as shown in Fig. 3b. First, noise was added to the simulated images of each detector segment based on the Poisson statistics. The noisy and noise-free images of each imaging method were then reconstructed. The assumed dose was equal to that of the experimental condition, as shown in Fig. 3a. The noise component images were then obtained by subtracting the noise-free images from their noisy counterparts, as shown in Fig. S2. The noise-component images were normalized using the contrast range of their corresponding noise-free image. The noise characteristics of different techniques were then compared, as shown in Fig. 3b.

**DFT calculations**

To relax the FAU twin-boundary structure and calculate the interface energy, we performed DFT calculations using the VASP code [40] with the rev-vdW-DF2 method [41], which is suitable for calculating zeolitic atomic structures and energies [42]. For the relaxation, we first relaxed the FAU bulk structure, whose data is available in the International Zeolite Association database [43]. The initial FAU twin boundary structure was then created by connecting the two FAU framework models with opposite stacking sequences. Finally, we obtained the relaxed FAU twin structure and calculated the interface energy $\Delta E_\mathrm{interface}$ as follows:

$$\Delta E_\mathrm{interface} = \frac{E_\mathrm{twin} - E_\mathrm{bulk}}{2A}, \tag{2}$$

where $A$ is the cross-sectional area of the interface, and $E_{\text{bulk}}$ and $E_{\text{twin}}$ are the total energies of the FAU bulk and twin boundary structures, respectively.

# References

[1] Y. Li, L. Li, J. Yu, Applications of Zeolites in Sustainable Chemistry, Chem. 3 (2017) 928–949. https://doi.org/10.1016/j.chempr.2017.10.009.

[2] Z. Wang, J. Yu, R. Xu, Needs and trends in rational synthesis of zeolitic materials, Chem. Soc. Rev. 41 (2012) 1729–1741. https://doi.org/10.1039/c1cs15150a.

[3] Y. Li, J. Yu, New Stories of Zeolite Structures: Their Descriptions, Determinations, Predictions, and Evaluations, Chem. Rev. 114 (2014) 7268–7316. https://doi.org/10.1021/cr500010r.

[4] S. Morishita, R. Ishikawa, Y. Kohno, H. Sawada, N. Shibata, Y. Ikuhara, Attainment of 40.5 pm spatial resolution using 300 kV scanning transmission electron microscope equipped with fifth-order aberration corrector, Microscopy. 67 (2018) 46–50. https://doi.org/10.1093/jmicro/dfx122.

[5] S.X. Wang, L.M. Wang, R.C. Ewing, Electron and ion irradiation of zeolites, J. Nucl. Mater. 278 (2000) 233–241. https://doi.org/10.1016/S0022-3115(99)00246-9.

[6] O. Ugurlu, J. Haus, A.A. Gunawan, M.G. Thomas, S. Maheshwari, M. Tsapatsis, K.A. Mkhoyan, Radiolysis to knock-on damage transition in zeolites under electron beam irradiation, Phys. Rev. B. 83 (2011) 1–4. https://doi.org/10.1103/PhysRevB.83.113408.

[7] J.W. Menter, The electron microscopy of crystal lattices, Adv. Phys. 7 (1958) 299–348. https://doi.org/10.1080/00018735800101287.

[8] L.A. Bursill, E.A. Lodge, J.M. Thomas, Zeolitic structures as revealed by high-resolution electron microscopy, Nature. 286 (1980) 111–113. https://doi.org/10.1038/286111a0.

[9] M. Haider, S. Uhlemann, E. Schwan, H. Rose, B. Kabius, K. Urban, Electron microscopy image enhanced, Nature. 392 (1998) 768–769. https://doi.org/10.1038/33823.

[10] C. Li, Q. Zhang, A. Mayoral, Ten Years of Aberration Corrected Electron Microscopy for Ordered Nanoporous Materials, ChemCatChem. 12 (2020) 1248–1269. https://doi.org/10.1002/cctc.201901861.

[11] T. Seki, Y. Ikuhara, N. Shibata, Toward quantitative electromagnetic field imaging by differential-phase-contrast scanning transmission electron microscopy, Microscopy. 70 (2021) 148–160. https://doi.org/10.1093/jmicro/dfaa065.

[12] H. Yang, I. MacLaren, L. Jones, G.T. Martinez, M. Simson, M. Huth, H. Ryll, H. Soltau, R. Sagawa, Y. Kondo, C. Ophus, P. Ercius, L. Jin, A. Kovács, P.D. Nellist, Electron ptychographic phase imaging of light elements in crystalline materials using Wigner distribution deconvolution, Ultramicroscopy. 180 (2017) 173–179. https://doi.org/10.1016/j.ultramic.2017.02.006.

[13] K. Ooe, T. Seki, Y. Ikuhara, N. Shibata, Ultra-high contrast STEM imaging for segmented/pixelated detectors by maximizing the signal-to-noise ratio, Ultramicroscopy. 220 (2021) 113133. https://doi.org/10.1016/j.ultramic.2020.113133.

[14] H. Rose, Nonstandard Imaging Methods in Electron Microscopy, Ultramicroscopy. 2 (1977) 251–267. https://doi.org/10.1016/S0304-3991(76)91538-2.

[15] T. Seki, Y. Ikuhara, N. Shibata, Theoretical framework of statistical noise in scanning transmission electron microscopy, Ultramicroscopy. 193 (2018) 118–125. https://doi.org/10.1016/j.ultramic.2018.06.014.

[16] S.D. Findlay, N. Shibata, H. Sawada, E. Okunishi, Y. Kondo, Y. Ikuhara, Dynamics of annular bright field imaging in scanning transmission electron microscopy, Ultramicroscopy. 110 (2010) 903–923. https://doi.org/10.1016/j.ultramic.2010.04.004.

[17] E. Yücelen, I. Lazić, E.G.T. Bosch, ., Sci. Rep. 8 (2018) 1–10. https://doi.org/10.1038/s41598-018-20377-2.

[18] N. Shibata, T. Seki, G. Sánchez-Santolino, S.D. Findlay, Y. Kohno, T. Matsumoto, R. Ishikawa, Y. Ikuhara, Electric field imaging of single atoms, Nat. Commun. 8 (2017) 15631. https://doi.org/10.1038/ncomms15631.

[19] Y. Sasaki, T. Suzuki, Y. Ikuhara, A. Saji, Direct Observation of Channel Structures in Zeolite Y and A with a Slow-Scan, Charge-Coupled-Device Camera, J. Am. Ceram. Soc. 78 (1995) 1411–1413. https://doi.org/10.1111/j.1151-2916.1995.tb08506.x.

[20] V. Alfredsson, T. Ohsuna, O. Terasaki, J.-O. Bovin, Investigation of the Surface Structure of the Zeolites FAU and EMT by High-Resolution Transmission Electron Microscopy, Angew. Chemie Int. Ed. English. 32 (1993) 1210–1213. https://doi.org/10.1002/anie.199312101.

[21] M. Shamzhy, M. Opanasenko, P. Concepción, A. Martínez, New trends in tailoring active sites in zeolite-based catalysts, Chem. Soc. Rev. 48 (2019) 1095–1149. https://doi.org/10.1039/c8cs00887f.

[22] M.-L.U. Cornelius, L. Price, S.A. Wells, L.F. Petrik, A. Sartbaeva, The steric influence of extra-framework cations on framework flexibility: an LTA case study, Zeitschrift Für Krist. - Cryst. Mater. 234 (2019) 461–468. https://doi.org/10.1515/zkri-2019-0016.

[23] A. Mayoral, Q. Zhang, Y. Zhou, P. Chen, Y. Ma, T. Monji, P. Losch, W. Schmidt, F. Schüth, H. Hirao, J. Yu, O. Terasaki, Direct Atomic-Level Imaging of Zeolites: Oxygen, Sodium in Na-LTA and Iron in Fe-MFI, Angew. Chemie Int. Ed. 59 (2020) 19510–19517. https://doi.org/10.1002/anie.202006122.

[24] Y. Zhang, D. Smith, J.E. Readman, A. Mayoral, Direct Imaging and Location of $Pb^{2+}$ and $K^+$ in EMT Framework-Type Zeolite, J. Phys. Chem. C. 125 (2021) 6461–6470. https://doi.org/10.1021/acs.jpcc.1c00550.

[25] K. Momma, F. Izumi, VESTA 3 for three-dimensional visualization of crystal, volumetric and morphology data, J. Appl. Crystallogr. 44 (2011) 1272–1276. https://doi.org/10.1107/S0021889811038970.

[26] H.G. Brown, N. Shibata, H. Sasaki, T.C. Petersen, D.M. Paganin, M.J. Morgan, S.D. Findlay, Measuring nanometre-scale electric fields in scanning transmission electron microscopy using segmented detectors, Ultramicroscopy. 182 (2017) 169–178. https://doi.org/10.1016/j.ultramic.2017.07.002.

[27] J.M. Newsam, M.M.J. Treacy, D.E.W. Vaughan, K.G. Strohmaier, W.J. Mortier, The structure of zeolite ZSM-20: Mixed cubic and hexagonal stackings of faujasite sheets, J. Chem. Soc. Chem. Commun. (1989) 493–495. https://doi.org/10.1039/C39890000493.

[28] Y. Sasaki, T. Suzuki, Y. Takamura, A. Saji, H. Saka, Structure analysis of the mesopore in dealuminated zeolite Y by high resolution TEM observation with slow scan CCD camera, J. Catal. 178 (1998) 94–100. https://doi.org/10.1006/jcat.1998.2130.

[29] J. Rittner, D. Seidman, 〈110〉 Symmetric Tilt Grain-Boundary Structures in Fcc Metals With Low Stacking-Fault Energies, Phys. Rev. B. 54 (1996) 6999–7015. https://doi.org/10.1103/PhysRevB.54.6999.

[30] Y. Sato, T. Mizoguchi, F. Oba, Y. Ikuhara, T. Yamamoto, Arrangement of multiple structural units in a [0001] Σ49 tilt grain boundary in ZnO, Phys. Rev. B. 72 (2005) 1–7. https://doi.org/10.1103/PhysRevB.72.064109.

[31] S. Fabris, S. Nufer, C. Elsässer, T. Gemming, Prismatic Σ3 (10-10) twin boundary in a-$Al_2O_3$ investigated by density functional theory and transmission electron microscopy, Phys. Rev. B. 66 (2002) 155415. https://doi.org/10.1103/PhysRevB.66.155415.

[32] N. Shibata, F. Oba, T. Yamamoto, Y. Ikuhara, Structure, energy and solute segregation behaviour of [110] symmetric tilt grain boundaries in yttria-stabilized cubic zirconia, Philos. Mag. 84 (2004) 2381–2415. https://doi.org/10.1080/14786430410001693463.

[33] M. V. Chubynsky, M.F. Thorpe, Self-organization and rigidity in network glasses, Curr. Opin. Solid State Mater. Sci. 5 (2001) 525–532. https://doi.org/10.1016/S1359-0286(02)00018-9.

[34] C.J. Dawson, R. Sanchez-Smith, P. Rez, M. O'Keeffe, M.M.J. Treacy, Ab initio calculations of the energy dependence of Si-O-Si angles in silica and Ge-O-Ge angles in germania crystalline systems, Chem. Mater. 26 (2014) 1523–1527. https://doi.org/10.1021/cm402814v.

[35] N. Shibata, Y. Kohno, S.D. Findlay, H. Sawada, Y. Kondo, Y. Ikuhara, New area detector for atomic-resolution scanning transmission electron microscopy, J. Electron Microsc. 59 (2010) 473–479. https://doi.org/10.1093/jmicro/dfq014.

[36] K. Yoshida, Y. Sasaki, Optimal accelerating voltage for HRTEM imaging of zeolite, Microscopy. 62 (2013) 369–375. https://doi.org/10.1093/jmicro/dfs087.

[37] R. Close, Z. Chen, N. Shibata, S.D. Findlay, Towards quantitative, atomic-resolution reconstruction of the electrostatic potential via differential phase contrast using electrons, Ultramicroscopy. 159 (2015) 124–137. https://doi.org/10.1016/j.ultramic.2015.09.002.

[38] L.J. Allen, A.J. D'Alfonso, S.D. Findlay, Modelling the inelastic scattering of fast electrons, Ultramicroscopy. 151 (2015) 11–22. https://doi.org/10.1016/j.ultramic.2014.10.011.

[39] J.M. Cowley, A.F. Moodie, The scattering of electrons by atoms and crystals. I. A new theoretical approach, Acta Crystallogr. 10 (1957) 609–619. https://doi.org/10.1107/S0365110X57002194.

[40] G. Kresse, J. Furthmüller, Efficient iterative schemes for ab initio total-energy calculations using a plane-wave basis set, Phys. Rev. B. 54 (1996) 11169–11186. https://doi.org/10.1103/PhysRevB.54.11169.

[41] I. Hamada, van der Waals density functional made accurate, Phys. Rev. B. 89 (2014) 121103. https://doi.org/10.1103/PhysRevB.89.121103.

[42] M. Fischer, W.J. Kim, M. Badawi, S. Lebègue, Benchmarking the performance of approximate van der Waals methods for the structural and energetic properties of $SiO_2$ and $AlPO_4$ frameworks, J. Chem. Phys. 150 (2019) 094102. https://doi.org/10.1063/1.5085394.

[43] C. Baerlocher, L.B. McCusker, Database of Zeolite Structures, http://www.iza-structure.org/databases/.

[44] K. Ooe, T. Seki, Y. Ikuhara, N. Shibata, High contrast STEM imaging for light elements by an annular segmented detector, Ultramicroscopy. 202 (2019) 148–155. https://doi.org/10.1016/j.ultramic.2019.04.011.

[45] K. Müller, F.F. Krause, A. Béché, M. Schowalter, V. Galioit, S. Löffler, J. Verbeeck, J. Zweck, P. Schattschneider, A. Rosenauer, Atomic electric fields revealed by a quantum mechanical approach to electron picodiffraction, Nat. Commun. 5 (2014) 5653. https://doi.org/10.1038/ncomms6653.

[46] I. Lazić, E.G.T. Bosch, S. Lazar, Phase contrast STEM for thin samples: Integrated differential phase contrast, Ultramicroscopy. 160 (2016) 265–280. https://doi.org/10.1016/j.ultramic.2015.10.011.

[47] L. Liu, D. Zhang, Y. Zhu, Y. Han, Bulk and local structures of metal–organic frameworks unravelled by high-resolution electron microscopy, Commun. Chem. 3 (2020) 1–14. https://doi.org/10.1038/s42004-020-00361-6.

[48] X. Li, I. Lazić, X. Huang, M. Wirix, L. Wang, Y. Deng, T. Niu, D. Wu, L. Yu, F. Sun, Imaging biological samples by integrated differential phase contrast (iDPC) STEM technique, J. Struct. Biol. 214 (2022) 107837. https://doi.org/10.1016/j.jsb.2022.107837.

# Acknowledgements

**Funding:** This work was supported by JSPS KAKENHI (Grant Numbers 20H05659, 20H00301, 20K15014), JST ERATO (Grant Number JPMJER2202), and a Grant-in-Aid for Specially Promoted Research "Atom-by-atom imaging of ion dynamics in nanostructures for materials innovation" (Grant Number 17H06094). K.O. acknowledges the support from the Grant-in-Aid for JSPS Research Fellow (Grant Numbers 19J23138, 22J01665). T.S. acknowledges the support from JST-PRESTO (Grant Number JPMJPR21AA) and the Kazato Research Foundation. A part of this study was conducted at the Research Hub for Advanced Nano Characterization, the University of Tokyo, with support from the Nanotechnology Platform (Project No. 12024046), MEXT, Japan.

**Author contributions:** K.O., T.S., and N.S. designed the study and wrote the paper. K.O. performed the STEM experiments, image simulations, and DFT calculations. K.Y. supported the STEM observation of zeolite samples and contributed to the discussions. Y.K. supported the development of the OBF STEM system and software. Y.I. contributed to the discussions and suggestions. N.S. and T.S. directed the entire study.

**Competing interests:** The authors declare the following financial interests/personal relationships which may be considered as potential conflict of interests: a part of the present authors are inventors on Japanese unexamined patent application publication filed by the University of Tokyo (No. 2021-077523).

**Data and materials availability:** All data necessary to evaluate the conclusions of this study are present in the paper and/or the Supplementary Materials. The data may be provided upon the request to the authors.

# Supplementary Materials

## Supplementary Text

**Dose efficiency evaluation based on the noise-normalized CTFs**

To compare the contrast transfer efficiency of different STEM techniques against the noise-level, we calculated the noise-normalized CTFs as shown in Fig. 1b. In this note, we show how to evaluate the theoretical dose efficiency of each technique by the noise-normalized CTFs.

Under the weak phase object approximation (WPOA), the STEM image intensity $I_{\text{STEM}}(\boldsymbol{R}_\text{p})$ is given as follows

$$I_{\text{STEM}}(\boldsymbol{R}_\text{p}) = d_0 + \mathcal{F}^{-1}[\sigma V(\boldsymbol{Q}_\text{p})\beta(\boldsymbol{Q}_\text{p})], \tag{S1}$$

where $d_0$, $\mathcal{F}^{-1}$, $\sigma$, $V(\boldsymbol{Q}_\text{p})$, and $\beta(\boldsymbol{Q}_\text{p})$ respectively indicate background intensity, the inverse Fourier transformation operator, the interaction parameter determined by the accelerating voltage of the electron beam, the Fourier component of the specimen projected potential, and the CTF. Assuming that the specimen projected potential in real space $v(\boldsymbol{R}_\text{p})$ is a delta function (i.e., $v(\boldsymbol{R}_\text{p}) \cong v_0 \delta(\boldsymbol{R}_\text{p})$ where $v_0$ is a constant), the STEM image intensity above the point scatterer ($\boldsymbol{R}_\text{p} = \boldsymbol{0}$) $I_{\text{STEM}}(\boldsymbol{0})$ is approximated as follows

$$I_{\text{STEM}}(\boldsymbol{0}) = d_0 + \sigma v_0 \int \beta(\boldsymbol{Q}_\text{p}) \, \text{d}\boldsymbol{Q}_\text{p}. \tag{S2}$$

This equation indicates that we can evaluate the contrast amplitude approximately by the second term, integration of CTF over the frequency domains, because the first term $d_0$ is the background. Thus, we compare the obtainable contrast against the noise-level among different STEM techniques by integrating the noise-normalized CTFs [13,44]. In the Poisson statistics, the signal-to-noise ratio (SNR) is proportional to the $\sqrt{\lambda}$, where $\lambda$ is the electron dose. Here, the integration value of noise-normalized CTFs can be regarded as a relative SNR between different techniques, and thus we can compare the dose-efficiency by the squared values of the CTF integration.

Table S1 shows the dose efficiency ratio calculated by the integration of the noise-normalized CTFs for OBF, iDPC, iCoM (integrated center-of-mass) [45,46], ABF, and conventional BF imaging methods, as discussed above. The iCoM meghod is a kind of iDPC imaging with a pixelated detector. Since we previously showed that OBF imaging can be extended to the pixelated detector [13], we calculated the CTF of OBF using the pixelated detector also. As for the segmented detector, the detector shape is the same as literatures for OBF [35] and iDPC [17], respectively. According to the calculated values shown in Table S1, OBF has approximately two-orders of magnitude higher dose efficiency than ABF theoretically. Furthermore, because OBF reconstructs the phase-contrast image in a theoretically optimized manner to obtain the highest SNR for each type of detector, the calculated dose efficiency of OBF is higher than the iDPC or iCoM techniques that use segmented or pixelated detectors, respectively. Since the iDPC technique is currently used for the STEM observation of beam-sensitive samples [47,48], the OBF observation should be able to reduce the irradiation dose more or obtain a higher spatial resolution on the same samples. Additionally, it should be noted that the OBF using a segmented detector obtains almost the same dose efficiency as iCoM that uses a pixelated detector. Although it is known that the pixelated detector can get significantly rich information about the sample, which could lead to higher dose-efficiency, this type of detector still needs longer dwell time while the recent technological progress improves the read-out speed. In the low-dose experiment, the operator must tune experimental conditions quickly under low SNR conditions. Thus, the capability of high-dose efficiency with a high-speed segmented detector is definitely helpful for beam-sensitive materials analysis.

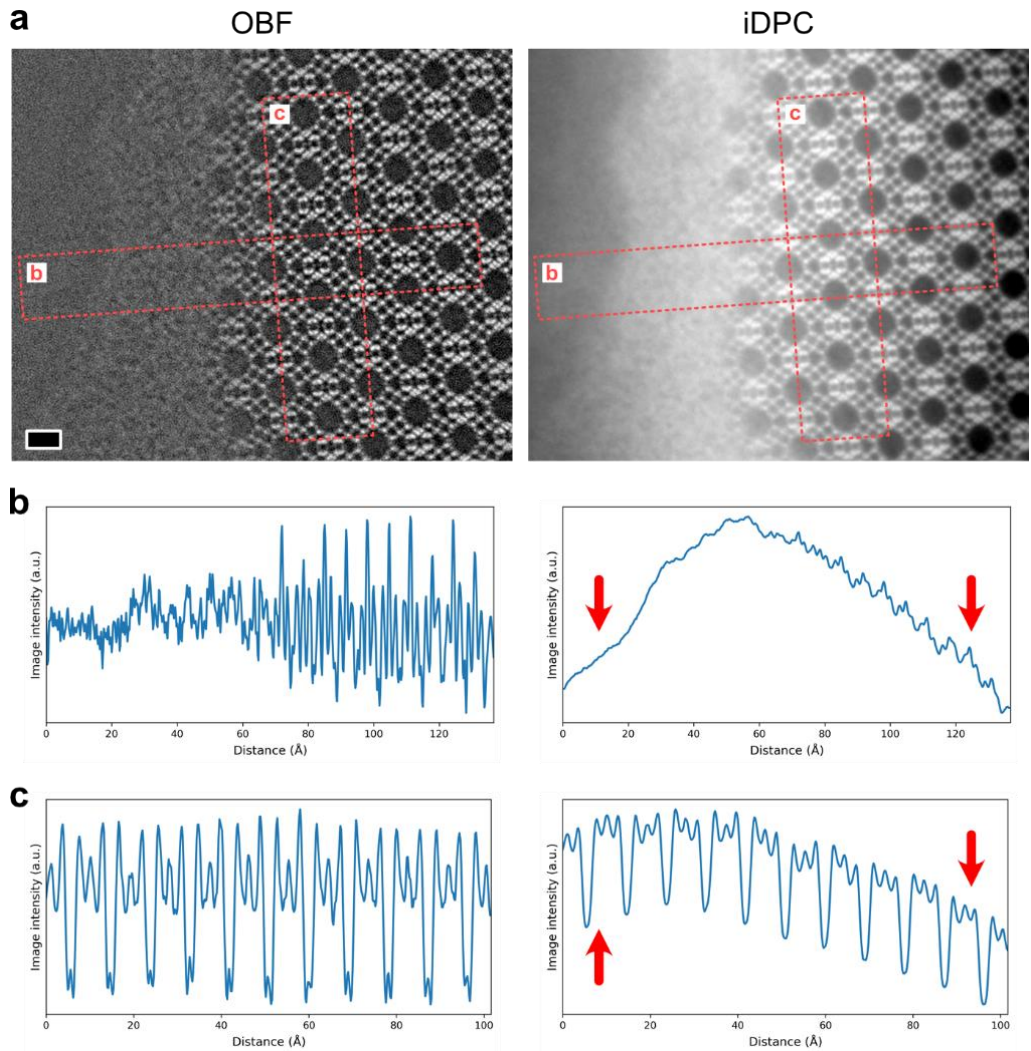

**Fig. S1. Comparison between OBF and iDPC images obtained from the same experimental dataset.** (a) OBF and iDPC images generated from the same dataset. The OBF image is the same as Fig. 2b. The intensity profiles taken along (b) [001] direction and (c) [1-10] direction respectively from the orange rectangles shown in the OBF and iDPC images. Since the observed sample is wedge-shaped and has an amorphous layer near the edge, the projected atomic potentials should be increased from left hand side (vacuum area) to right hand side (thicker sample area) along the direction shown in (b). On the other hand, along the direction shown in (c), the thickness is almost uniform and corresponding image contrast should also be uniform. These are the case for the OBF image, but the iDPC image has strong intensity fluctuations (indicated by orange arrows) as discussed in Fig. 3b.

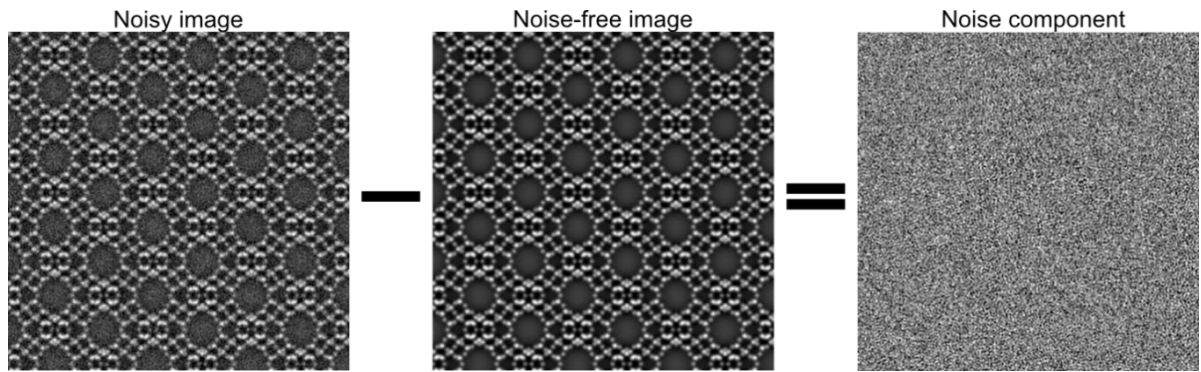

**Fig. S2. Schematic of noise evaluation technique.** Schematic illustration of noise evaluation method shown in Fig. 3b. By combining the noise-free image and noisy image, noise characteristics against the contrast range can be calculated.

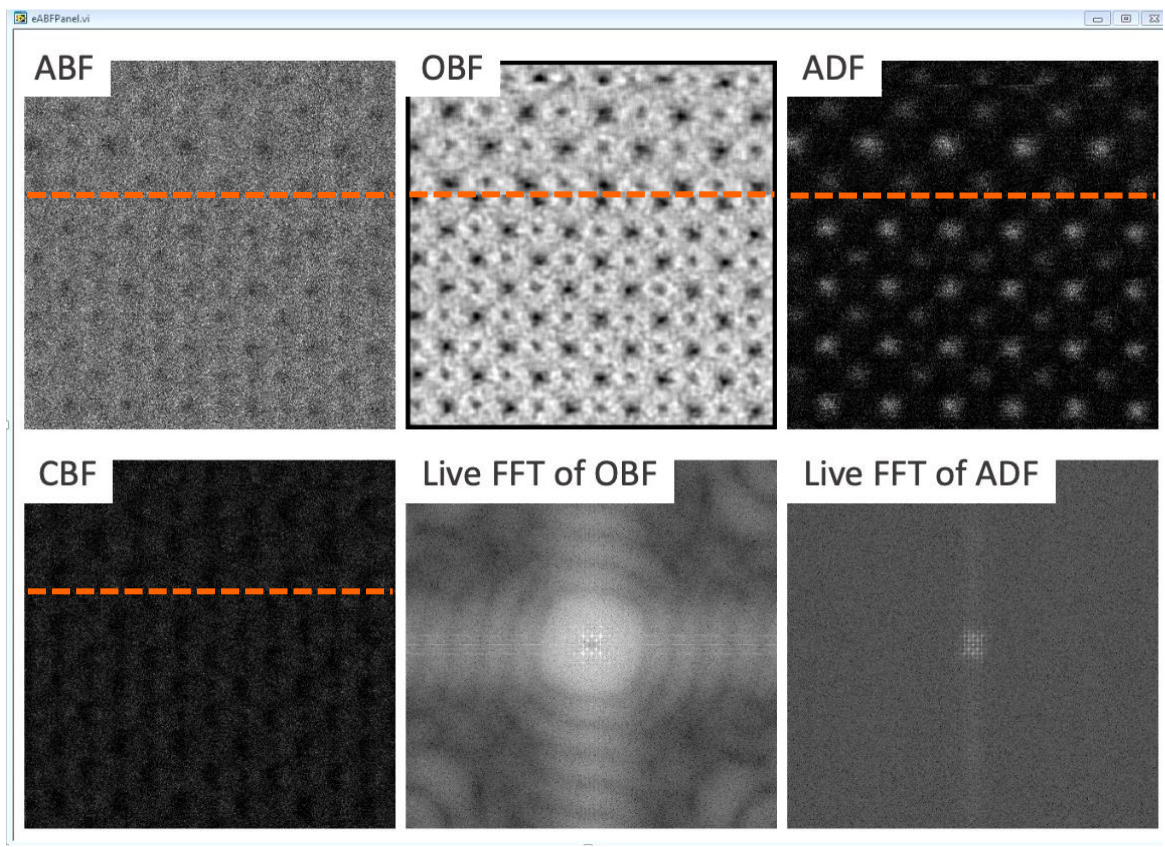

**Fig. S3. Display of live OBF imaging system.** Captured image of Movie S1 and its description. Movie S1 shows the real-time atomic-resolution OBF imaging of SrTiO$_3$ [001] under a low-dose condition. The dwell time is 10 µs, and the image is sampled with 512x512 pixels. In the upper row, the left panel, center panel, and the right panel show ABF, OBF, and annular dark-field (ADF) images, respectively. In the lower row, the left panel shows a center bright-field (CBF) image, and the center and right panel shows Fourier transformed OBF and ADF images, respectively. The images, including OBF, are synchronized with STEM probe scans and updated in real-time. The updated area in this capture is highlighted with a dotted line in each image. Movie S1 also shows the area scan mode, where the only selected area inside the image is scanned and the frame rate is increased for tuning aberrations such as defocus.

**Table S1. Comparison of dose efficiency of different STEM methods.** Dose efficiency ratio among different STEM imaging techniques based on the noise-normalized CTF calculations. The values are normalized such that the dose-efficiency of ABF becomes one.

| Detector type | Method | Dose-efficiency ratio |
|---|---|---|
| Pixelated | OBF | 110 |
| | iDPC (iCoM) | 69.9 |
| Segmented | OBF | 66.5 |
| | iDPC | 53.8 |
| Annular (conventional) | BF | 1.02 |
| | ABF | 1 |

**Movie S1.**

Live atomic-resolution OBF imaging of SrTiO$_3$ [001] using the real-time OBF system.